\documentclass[aps,prd,twocolumn,nofootinbib,superscriptaddress,longbibliography,floatfix]{revtex4-2}
\usepackage{amsmath,amssymb,bm}
\usepackage{graphicx}
\usepackage{booktabs}
\usepackage{xcolor}
\usepackage{hyperref}
\usepackage{microtype}

\hypersetup{
  colorlinks=true,
  linkcolor=blue!45!black,
  citecolor=blue!45!black,
  urlcolor=blue!45!black,
  pdftitle={Atomic clocks, free-fall tests, and the relativistic galaxy dipole as probes of quintessence coupled to photons and dark matter},
  pdfauthor={Alessandro Melchiorri},
  pdfsubject={Atomic clocks, local and cosmological gravitational redshift, and scalar interacting dark energy},
  pdfkeywords={atomic clocks, gravitational redshift, interacting dark energy, quintessence, varying constants}
}

\newcommand{\mpl}{M_{\rm Pl}}
\newcommand{\dd}{\mathrm{d}}
\newcommand{\Rvd}{\mathcal{R}_{vd}}

\begin{document}

\preprint{Accepted for publication in Phys. Rev. D}

\title{Atomic clocks, free-fall tests, and the relativistic galaxy dipole\texorpdfstring{\\}{ }as probes of quintessence coupled to photons and dark matter}

\author{Alessandro Melchiorri}
\email{alessandro.melchiorri@uniroma1.it}
\affiliation{Physics Department, Sapienza University of Rome, Piazzale Aldo Moro 5, 00185 Rome, Italy}

\date{\today}

\begin{abstract}
We consider a light quintessence field that couples to photons through the
gauge kinetic function and to cold dark matter through a conformal factor, with
baryons minimally coupled, and ask which combination of its two couplings each
available observable constrains. Three observables are compared: the secular
drift of atomic-clock frequency ratios, the differential free fall of test masses,
and the antisymmetric dipole of the cross-correlation function of two galaxy
populations, a relativistic effect in a homogeneous background that arises when
the two populations respond differently to the gravitational potential. A clock
measures the coupling of the scalar to electromagnetism, multiplied by the
present rate of change of the field. The background energy transfer measures the
coupling of the scalar to cold dark matter, multiplied by the same rate of
change; the dipole instead constrains a force-, range- and tracer-dependent
combination that is quadratic in that coupling. We give, for each observable, the combination of
parameters it constrains, and the assumptions needed to convert one combination
into another. For a single unscreened scalar with linear conformal couplings, the
rate of change of the field cancels between clock drift and background transfer,
so a null clock result and a measured dark-sector transfer rate together bound
the ratio of the two couplings without requiring the field trajectory to be
known. Current Yb$^+$ data put that ratio below $2\times10^{-7}$ if the transfer is
five percent per Hubble time. Treating the two clock channels jointly, we find
that imposing a grand-unified relation between the photon and mass-ratio
couplings tightens the photon limit by less than ten percent, and that the
mass-ratio channel reaches parity only for a response factor near $10^2$. 
\end{abstract}

\maketitle

\section{Introduction}
\label{sec:intro}

Measurements of distances, clustering and lensing constrain the late-time
expansion history of the Universe at the percent level \cite{Planck2020,DESI2025DR2}. They can
test whether the dark-energy equation of state differs from $w=-1$. They cannot
say what causes such a difference. A rolling scalar field
\cite{Wetterich1995,Carroll1998}, a modification of gravity on large scales, and
an interaction between dark energy and cold dark matter \cite{Amendola2000} can
all produce the same expansion history. Growth and lensing data reduce the
number of options, but within the model class studied in Ref.~\cite{Castello2024PRD}
they do not separate a fifth force acting only on dark matter from a
modification of gravity that acts on everything. To separate these cases one has to know what the field
couples to.

Couplings can be measured in the laboratory as well as in cosmology. In the
laboratory, the ratio of two atomic clock frequencies depends on dimensionless
constants such as the fine-structure constant $\alpha$ and the
proton-to-electron mass ratio $\mu$. Comparing clocks over several years tests
whether these constants drift. Comparing them while the laboratory samples a
changing gravitational potential tests local position invariance (LPI). In
cosmology, a scalar coupled to dark matter exchanges energy with it and exerts
an extra force on it, which changes the growth of structure and the velocities
of galaxies.

The same term is used for two different measurements. In a clock experiment,
gravitational redshift is the difference in rate between two clocks at
different potentials. In large-scale structure, it is a relativistic
contribution to the observed redshift of a galaxy, which produces an
antisymmetric dipole in the cross-correlation of two galaxy populations
\cite{Bonvin2014,Castello2024JCAP,Castello2024PRD}. The two effects have a
common relativistic origin, but different sources, different systematic errors,
and different scalar couplings.

Because the couplings are different, a result of one kind cannot be converted
into the other without additional assumptions. Interacting dark energy shows
this clearly. An exchange term $Q$ added to the continuity equations can be
constrained by distance and growth data, and many forms have been used:
$Q\propto H\rho_{\rm DE}$, $Q\propto H\rho_c$, nonlinear combinations of the
densities, sign-changing interactions, momentum exchange, diffusion, and
covariant conformal or disformal couplings
\cite{Amendola2000,Gleyzes2015,CosmoVerse2025}. These are not equivalent to one
another. A coefficient fitted in a continuity equation is not a coupling
constant of a field, and on its own it predicts nothing about an atomic
transition. The opposite conversion is also common. A bound on
$\dot\alpha/\alpha$ is often quoted as a bound on $w+1$, but a clock measures a
coupling multiplied by a field velocity, so the result depends on the assumed
coupling as much as on the equation of state
\cite{Uzan2025,Martins2015,Martins2017}.

The model we use is the minimal one in which both signals exist: a canonical
scalar with a quintessence potential, coupled to photons and to cold dark matter,
with baryons minimally coupled. Its dark-matter sector is the two-coupling
scalar-tensor class studied for nucleosynthesis in Ref.~\cite{Coc2009} and,
with a constant coupling and a cosmological constant, as the $\Lambda\beta$CDM
model of Refs.~\cite{PitrouUzan2024,UzanPitrou2024}; the present work extends
that class by a photon coupling and a rolling potential, and all quantitative
statements below are made within it. The full field equations are collected in
Appendix~\ref{app:fieldeqs}.

Clock--quintessence relations and their equivalence-principle consequences have
long been studied \cite{Martins2015,Martins2017}. Particularly relevant to the
present work is Ref.~\cite{Archidiacono2022}, which used the same clock-drift
measurements to constrain dilaton-like photon and QCD couplings in pure-fifth-force
and coupled-dark-energy scenarios, compared those bounds with
equivalence-principle tests, and translated them into constraints on the
dark-force strength in explicit microscopic models. In our conventions its
electromagnetic coefficient satisfies $|\zeta_\gamma|=|d_e|/\sqrt{2}$, so its
clock bound has the same $[\Omega_s^0(1+w_s^0)]^{-1/2}$ dependence as
Eq.~\eqref{eq:zetabound}. The small difference in the numerical coefficient
follows mainly from the confidence-interval prescription: Ref.~\cite{Archidiacono2022}
neglects the measured central value, whereas we use the largest absolute endpoint
of the two-sided 95\% interval.

Our emphasis is therefore not the clock-only quintessence scaling or the
qualitative clock-versus-equivalence-principle ordering. We instead keep the
photon and CDM charges independent and organize the comparison at the observable
level: clock drift constrains $\zeta_\gamma\varphi'$, background energy transfer
constrains $\beta_c\varphi'$, while the relativistic galaxy dipole constrains a
force-, range-, and tracer-dependent combination beginning at order $\beta_c^2$.
This separation yields the trajectory-independent identity
$\Theta_\alpha=(\zeta_\gamma/\beta_c)\xi_c$ only when the background transfer
rate is independently determined, and explains why the same cancellation cannot
be inferred from the dipole or from a phenomenological interaction coefficient
alone. Keeping the visible charges independent also lets us treat the two clock
channels jointly rather than one at a time, which quantifies how little the
$\mu$ channel currently contributes even in a model that correlates it with
$\alpha$. We work in a minimal model with one scalar coupled to photons and to
cold dark matter, evaluate the relevant parameter combinations with published
data, and list the conditions under which they may be related. Four questions organize the argument:

\begin{enumerate}
 \item What combination of field dynamics and visible-sector coupling do clocks
 measure?
 \item What dark-sector combination does the cosmological
 gravitational-redshift dipole measure?
 \item Under what assumptions can clock drift and an independently determined
 background transfer rate be combined, and what complementary information does
 the dipole add?
 \item Which dark-energy and IDE models make no prediction for this comparison
 at all?
\end{enumerate}

The answer to the third question is the ratio $\Rvd=|\zeta_\gamma/\beta_c|$ of
the photon coupling to the dark-matter coupling. This ratio stays meaningful
even when each coupling separately is degenerate with the field trajectory.
Section~\ref{sec:framework} sets up the model and the sign conventions.
Section~\ref{sec:clocks} derives what a clock ratio measures, compares the
drift, potential-modulation and free-fall channels once the electromagnetic
charges are assigned, and gives the joint region for the two visible couplings.
Section~\ref{sec:lss} does the same for the cosmological dipole,
Sec.~\ref{sec:joint} combines the two and lists the six conditions to check
before using the hierarchy relation, and Sec.~\ref{sec:taxonomy} classifies the
models for which the combination applies. Table~\ref{tab:models} states which
probe applies to which model class. All limits below are recalculations of published
measurements with the charge conventions written out; forecasts are either
rescalings of present sensitivity or results taken from the cited survey
studies.

We use $c=\hbar=1$, metric signature $(-,+,+,+)$, and the reduced Planck mass
$\mpl=(8\pi G)^{-1/2}$. The dimensional field is $\phi$, and its dimensionless
roll is written $\varphi'\equiv\dd(\phi/\mpl)/\dd\ln a$ throughout.

\section{A two-sector scalar framework}
\label{sec:framework}

\subsection{Action and couplings}

The model is defined by the Einstein-frame action
\begin{align}
 S={}&\int \dd^4x\sqrt{-g}\left[\frac{\mpl^2}{2}R
 -\frac12(\partial\phi)^2-V(\phi)\right]
 \nonumber\\
 &+S_{\rm SM}\!\left[g_{\mu\nu},\psi_{\rm SM};\,\alpha(\phi),m_e(\phi),m_q(\phi),\Lambda_{\rm QCD}(\phi)\right]
 \nonumber\\
 &+S_c[A_c^2(\phi)g_{\mu\nu},\psi_c] .
\label{eq:action}
\end{align}
Three choices are built into it and are kept fixed throughout.

First, the Standard-Model fields are minimally coupled to $g_{\mu\nu}$: there is
no baryonic conformal factor, $A_b\equiv1$. This fixes what we mean by the
Einstein frame: it is the frame in which the Planck mass and the non-electromagnetic
part of baryonic masses are constant, and therefore the frame in which laboratory
rods, clocks, and the Cavendish constant $G_N=(8\pi\mpl^2)^{-1}$ are defined. A
universal $A_b(\phi)$ would rescale all atomic frequencies together and could be
absorbed into a redefinition of that frame; it cannot be detected by a frequency
ratio and is not the object of this paper. The Jordan frame of the dark-matter
sector, $\tilde g_{\mu\nu}=A_c^2 g_{\mu\nu}$, is the one in which CDM particles
follow geodesics; it differs from the baryonic one, which is what makes the
theory non-universal. With $A_b=1$ the dark-matter sector is the two-coupling
scalar-tensor class of Refs.~\cite{Coc2009,PitrouUzan2024,UzanPitrou2024}, and
the constraints on that class from primordial nucleosynthesis \cite{Coc2009}
apply here unchanged.

Second, the scalar enters the visible sector only through the dimensionless
Standard-Model parameters. In the electromagnetic sector this is the gauge kinetic
function, $-\tfrac14 B_F(\phi)F_{\mu\nu}F^{\mu\nu}$, so that
\begin{equation}
 \alpha(\phi)=\frac{\alpha_0}{B_F(\phi)}
\label{eq:alphaBF}
\end{equation}
exactly, whatever the form of $B_F$; the linearisation $B_F=1-\zeta_\gamma(\phi-\phi_0)/\mpl+\cdots$ serves only to define the coupling. Our benchmark throughout is the
\emph{photon-only} case in which $B_F$ is the only $\phi$-dependent visible
parameter. The general case, in which $\Lambda_{\rm QCD}$, $m_q$, and $m_e$ also
depend on $\phi$, is the dilaton model of Damour and Donoghue
\cite{DamourDonoghue2010}; we invoke it only where explicitly stated, namely for the
mass-ratio channel of Sec.~\ref{sec:clocks}\,A and~E and for the alternative
equivalence-principle recasts of Sec.~\ref{sec:clocks}\,D. Cosmological redshift
is unaffected by either choice: photons propagate on $g_{\mu\nu}$, so the
frequency of a photon scales as $a^{-1}$ and $1+z=a_0/a$ for a comoving emitter. What
$B_F$ changes is the \emph{emitted} frequency of an atomic line, so that a line
observed at redshift $z$ satisfies
$\nu_{\rm obs}=\nu_{\rm lab}(1+z)^{-1}[1+K_\alpha\,\delta\ln\alpha(z)]$, with
$K_\alpha$ the sensitivity of that transition; this is the effect exploited by
quasar absorption-line searches and, at $z=0$, by clock comparisons.

Third, the dark-matter conformal factor $A_c(\phi)$ is the only coupling of the
scalar to the dark sector. Near the present value $\phi_0$ we define
\begin{align}
 \zeta_\gamma &\equiv
 \left.\frac{\dd\ln\alpha}{\dd(\phi/\mpl)}\right|_0,
 &
 \beta_c &\equiv
 \left.\frac{\dd\ln A_c}{\dd(\phi/\mpl)}\right|_0,
\label{eq:couplings}
\end{align}
and, for the general dilaton case, $\zeta_\mu\equiv\dd\ln\mu/\dd(\phi/\mpl)|_0$
with $\mu\equiv m_p/m_e$. Even in the photon-only case the baryons are not
exactly neutral: the electromagnetic binding energy of a nucleus depends on
$\alpha$, so a body $A$ acquires the induced charge
$\beta_A^{\rm(em)}=\zeta_\gamma\,\partial\ln m_A/\partial\ln\alpha$, of order
$10^{-3}\zeta_\gamma$ for ordinary matter \cite{DamourDonoghue2010}. This is the
charge that equivalence-principle tests measure (Sec.~\ref{sec:clocks}\,D); for the
values of $\zeta_\gamma$ allowed by the clock bounds it is below $10^{-10}$ and is
dropped from the background equations. Likewise, in the photon-only case the
proton-to-electron mass ratio varies only through the electromagnetic
self-energy of the proton, $\zeta_\mu=Q_e^{p}\,\zeta_\gamma$ with
$Q_e^{p}\equiv\partial\ln m_p/\partial\ln\alpha\simeq7\times10^{-4}$, so the
$\mu$ channel carries no independent information; $\zeta_\mu$ is a free parameter
only in the general dilaton case.

The action is a benchmark, not a claim of universality. Kinetic, disformal,
derivative, multi-field, and screened models fall outside it and can alter every
mapping derived below; Sec.~\ref{sec:taxonomy} lists them and states which of the
present assumptions each one replaces. The advantage of Eq.~\eqref{eq:action} is
that it separates three logically different ingredients: the cosmological
trajectory $\phi(a)$, the visible coupling $\zeta_\gamma$, and the dark coupling
$\beta_c$.

\subsection{Background exchange and dark force}

In a flat FLRW background the field equations derived from Eq.~\eqref{eq:action}
are given in Appendix~\ref{app:fieldeqs}. The ones that carry the dark coupling are
\begin{align}
 \ddot\phi+3H\dot\phi+V_{,\phi} &=-\frac{\beta_c}{\mpl}\,\rho_c,\nonumber\\
 \dot\rho_c+3H\rho_c &= Q_c\equiv\beta_c\rho_c\frac{\dot\phi}{\mpl},
 \label{eq:conservation}\\
 \dot\rho_\phi+3H(\rho_\phi+p_\phi) &=-Q_c,\nonumber
\end{align}
with $\rho_\phi=\tfrac12\dot\phi^2+V$ and $p_\phi=\tfrac12\dot\phi^2-V$, while
baryons and radiation obey the standard conservation equations. The third line
follows from the first two and states that the energy leaving the CDM sector
enters the scalar one; the sum of all sectors is conserved. Equation~\eqref{eq:conservation}
defines the sign of $Q_c$ used throughout: with $\beta_c\dot\phi>0$ the
CDM mass grows and energy flows from the field into dark matter. The second line
integrates to
\begin{equation}
 \rho_c(a)=\rho_{c0}\,\frac{A_c(\phi(a))}{A_c(\phi_0)}\,a^{-3},
\label{eq:rhoc}
\end{equation}
which is exact for any $A_c$ \cite{Coc2009,UzanPitrou2024}: the CDM number
density dilutes as $a^{-3}$ and the CDM mass follows $A_c$. It is convenient to
introduce
\begin{equation}
\begin{aligned}
 \varphi&\equiv\frac{\phi}{\mpl},\qquad
 \varphi'\equiv\frac{\dd\varphi}{\dd\ln a}
 =\frac{\dot\phi}{H\mpl},\\
 \xi_c&\equiv\frac{Q_c}{H\rho_c}=\beta_c\varphi'.
\end{aligned}
\label{eq:xic}
\end{equation}
The quantity $\xi_c$ is the fractional CDM energy-transfer rate per Hubble time in our sign convention. It is linear in $\dot\phi$ because it is the logarithmic rate of change of the CDM mass, $\xi_c=\dd\ln A_c/\dd\ln a$; this should not be confused with the scalar's own energy density, whose kinetic part is quadratic in $\dot\phi$.

The same coupling modifies the CDM Euler equation, Eq.~\eqref{eq:eulerc} below and
Appendix~\ref{app:fieldeqs}. In the quasistatic sub-horizon limit the scalar
perturbation sourced by a CDM overdensity adds to the Newtonian force, and the
Poisson equation for CDM perturbations sourced by CDM acquires an effective
coupling $G_{cc}(k,a)$ in place of the Cavendish constant $G_N$,
\begin{equation}
 \frac{G_{cc}(k,a)}{G_N}-1
 = 2\beta_c^2\frac{k^2}{k^2+a^2m_\phi^2},
\label{eq:fifthforce}
\end{equation}
where $k$ is the comoving wavenumber and $m_\phi$ the effective scalar mass on the
background,
\begin{equation}
 m_\phi^2=V_{,\phi\phi}+\frac{\beta_c^2\rho_c}{\mpl^2}
\label{eq:mphi}
\end{equation}
for $A_c=\exp(\beta_c\phi/\mpl)$. The factor of $2$ follows from this
normalization and moves under others, which is why Appendix~\ref{app:conventions}
insists that couplings be compared through Eq.~\eqref{eq:fifthforce} rather than
by matching symbols. For a quintessence potential $V_{,\phi\phi}\sim H^2$, and
with $\beta_c\lesssim0.1$ the second term is of the same order, so $m_\phi\sim H$ and
the $k$-dependent factor equals unity on all sub-horizon scales: within
Eq.~\eqref{eq:action} the dark force is long range. The factor is retained because
the finite-range models of Sec.~\ref{sec:taxonomy} are the ones in which it matters.
Baryons do not feel this force beyond the induced charge $\beta_b^{\rm(em)}$, so
$G_N$ measured with baryonic test masses is the bare constant.

Equation~\eqref{eq:fifthforce} makes one distinction explicit that the rest of the
paper relies on. The energy transfer $\xi_c=\beta_c\varphi'$ is linear in $\beta_c$,
because it is one charge multiplied by the velocity of the field. The force is
quadratic, because it is the charge of the source multiplied by the charge of
the test particle, and both are $\beta_c$. A measurement of the force therefore
does not determine $\xi_c$, and a measurement of $\xi_c$ does not determine the
force, unless $\varphi'$ is known independently. Finally, in the finite-range
models a Compton wavelength shorter than survey scales suppresses the force
without affecting the homogeneous energy exchange.

\subsection{Phenomenological versus microphysical IDE}

The frequently used parameterization
\begin{equation}
 Q_c=\xi_{\rm DE}H\rho_{\rm DE}
\label{eq:Qphen}
\end{equation}
implies, at the background level,
\begin{equation}
 \xi_c=\xi_{\rm DE}\frac{\rho_{\rm DE}}{\rho_c}
 =\xi_{\rm DE}\frac{\Omega_{\rm DE}}{\Omega_c}.
\label{eq:ximap}
\end{equation}
Here $\rho_{\rm DE}$ is a fluid; in the microphysical completion of Eq.~\eqref{eq:action} it is identified with $\rho_\phi=\tfrac12\dot\phi^2+V$, the field being the dark energy, and $1+w_\phi=\dot\phi^2/\rho_\phi$. This identity holds instantaneously at each redshift but does not specify $\beta_c$, $\varphi'$, the scalar mass, perturbations, or the coupling to photons. In particular, a fit that takes $\xi_{\rm DE}$ to be constant over the full expansion history is not equivalent to a model with both constant $\beta_c$ and arbitrary canonical $\varphi'(a)$: the latter gives $\xi_c(a)=\beta_c\varphi'(a)$, which follows the field velocity and switches off as the scalar freezes, whereas Eq.~\eqref{eq:ximap} gives $\xi_c(a)=\xi_{\rm DE}\Omega_{\rm DE}(a)/\Omega_c(a)$, which grows monotonically towards late times whatever the field is doing. The two agree only if the trajectory is chosen to make them agree. A phenomenological constraint on $\xi_{\rm DE}$ becomes a clock prediction only after one asserts Eq.~\eqref{eq:conservation}, identifies the field in Eqs.~\eqref{eq:conservation} and \eqref{eq:couplings}, and fits the same redshift-dependent model.

\section{What atomic clocks measure}
\label{sec:clocks}

\subsection{Frequency ratios and secular drift}

An observable clock comparison is a dimensionless frequency ratio $R_{AB}=\nu_A/\nu_B$. Its response to slowly varying constants is
\begin{equation}
 \delta\ln R_{AB}=K_\alpha^{AB}\delta\ln\alpha
 +K_\mu^{AB}\delta\ln\mu
 +K_q^{AB}\delta\ln X_q+\cdots,
\label{eq:clockresponse}
\end{equation}
where the $K$ coefficients are calculated atomic or molecular sensitivities and $X_q$ denotes a dimensionless light-quark/QCD combination \cite{Uzan2025}. This is the general expression; in the photon-only case of Eq.~\eqref{eq:action} it reduces to $\delta\ln R_{AB}=K_\alpha^{AB}\,\delta\ln\alpha$, since $X_q$ is constant and $\delta\ln\mu=Q_e^p\,\delta\ln\alpha$ is negligible. Multiple, compositionally distinct clocks are needed to separate the terms in the general case. A frequency standard alone does not measure an absolute variation of a dimensional constant.

It is convenient to measure the drift in Hubble units. For the electromagnetic coupling in Eq.~\eqref{eq:couplings} we define
\begin{equation}
 \Theta_\alpha\equiv\frac{1}{H}\frac{\dot\alpha}{\alpha}
 =\zeta_\gamma\varphi',
\label{eq:theta}
\end{equation}
so that a pair of clocks compared over a time $\Delta t\ll H^{-1}$ records $\delta\ln R_{AB}=K_\alpha^{AB}\,H\Theta_\alpha\,\Delta t$. A null drift can mean a small coupling, a nearly stationary field, or both, and a clock alone cannot tell which. As a canonical field approaches $w=-1$ the factor $\varphi'$ goes to zero, so the bound on $\zeta_\gamma$ that a fixed drift limit implies weakens without bound.

Equation~\eqref{eq:theta} also assumes that the scalar time derivative sampled by the laboratory follows the homogeneous cosmological roll. A screened, virialized, or environmentally pinned local solution replaces $\varphi'$ by the corresponding local derivative and invalidates the direct clock--cosmology mapping.

In the general dilaton case the mass-ratio channel is parallel,
\begin{equation}
 \Theta_\mu\equiv\frac{1}{H}\frac{\dot\mu}{\mu}
 =\zeta_\mu\varphi',
\label{eq:thetamu}
\end{equation}
with $\zeta_\mu$ defined in Sec.~\ref{sec:framework}\,A. In the photon-only case $\zeta_\mu=Q_e^p\zeta_\gamma\simeq7\times10^{-4}\zeta_\gamma$ and the channel adds nothing. It becomes an independent probe only when the scalar also couples to the gluon or quark-mass operators, and the unification benchmark $\zeta_\mu=R\,\zeta_\gamma$ with $|R|=30$--$50$ used below \cite{Martins2017} refers to that case.

The extended E2/E3 comparison in $^{171}{\rm Yb}^+$ gives the current secular $\alpha$ result \cite{Filzinger2023}, while the cesium-referenced analysis supplies the $\mu$ drift \cite{Lange2021}:
\begin{align}
 \frac{\dot\alpha}{\alpha}&=(1.8\pm2.5)\times10^{-19}\ {\rm yr}^{-1},
\label{eq:clockdrift}\\
 \frac{\dot\mu}{\mu}&=(-8\pm36)\times10^{-18}\ {\rm yr}^{-1}.
\end{align}
Earlier optical comparisons established the long-baseline method \cite{Rosenband2008}. Adopting $H_0=67.4\ {\rm km\,s^{-1}\,Mpc^{-1}}=6.893\times10^{-11}\ {\rm yr}^{-1}$, the absolute upper endpoint of the two-sided Gaussian 95\% interval, $|\bar x|+1.96\sigma_x$, gives
\begin{equation}
 |\Theta_{\alpha0}|_{95}<9.72\times10^{-9}.
\label{eq:theta95}
\end{equation}
The corresponding $\mu$ endpoint is
\begin{equation}
 |\Theta_{\mu0}|_{95}<1.14\times10^{-6}.
\label{eq:thetamu95}
\end{equation}
The raw dimensionless $\mu$ limit is now $117$ times weaker. A unification benchmark $\Delta\mu/\mu=R\,\Delta\alpha/\alpha$ with $|R|=30$--$50$ makes it equivalent to $|\Theta_\alpha|<(2.28$--$3.80)\times10^{-8}$, still $2.3$--$3.9$ times weaker than Eq.~\eqref{eq:theta95} \cite{Martins2017}. The channel remains essential because larger QCD response factors or future molecular sensitivity can reverse that ordering, but this is a conditional correlated-coupling benchmark, not a model-independent conversion. Since $\Theta_X\propto H_0^{-1}$, using $H_0=73\ {\rm km\,s^{-1}\,Mpc^{-1}}$ would tighten the quoted dimensionless bounds by $7.7\%$.

\subsection{Potential modulation and local gravitational redshift}

The same clock record can be searched for modulation as the terrestrial laboratory samples the Sun's potential. We define $U=+GM/r$ as the positive potential depth and $\Delta U$ as the value at the later sample minus the earlier one. Writing
\begin{equation}
 \delta\ln X=k_X\frac{\Delta U}{c^2},
\label{eq:kx}
\end{equation}
the updated Yb$^+$ analysis found \cite{Filzinger2023}, while the $\mu$ coefficient is from Ref.~\cite{Lange2021},
\begin{align}
 k_\alpha&=(-2.4\pm3.0)\times10^{-9},
 &k_\mu&=(7\pm45)\times10^{-8}.
\label{eq:kresults}
\end{align}
For a light, unscreened scalar sourced by a body whose effective scalar charge is $\beta_s$, defined as in Eq.~\eqref{eq:couplings} with $A$ the mass function of the source and evaluated for the Sun in Sec.~\ref{sec:clocks}\,D, the weak-field solution in the conventions of Eq.~\eqref{eq:action} is $\Delta\phi/\mpl\simeq2\beta_s\Delta U/c^2$. Hence
\begin{equation}
 k_\alpha\simeq2\zeta_\gamma\beta_s.
\label{eq:kalphamap}
\end{equation}
The absolute 95\% endpoint of Eq.~\eqref{eq:kresults} gives $|k_\alpha|<8.28\times10^{-9}$, or $|\zeta_\gamma\beta_s|<4.14\times10^{-9}$ under these assumptions. Reversing the sign convention for $U$ reverses Eq.~\eqref{eq:kalphamap} but leaves this limit unchanged. The result constrains a product involving the scalar charge of the source; it is not equivalent to Eq.~\eqref{eq:theta95} and need not involve $\beta_c$.

Traditional redshift experiments instead write
\begin{equation}
 \frac{\Delta\nu}{\nu}=(1+\alpha_{\rm LPI})\frac{\Delta U}{c^2},
\label{eq:lpiredshift}
\end{equation}
after modelling kinematic and instrumental terms. The eccentric Galileo satellites yield $\alpha_{\rm LPI}=(0.19\pm2.48)\times10^{-5}$ in one analysis \cite{Delva2018}, with an independent analysis reaching comparable precision \cite{Herrmann2018}; these improved the classic Gravity Probe A result \cite{Vessot1980}. The Hafele--Keating flights were foundational demonstrations of gravitational and kinematic time dilation \cite{HafeleKeating1972a,HafeleKeating1972b}, but they are not competitive constraints on scalar dark energy.

Optical clocks now resolve the standard gravitational redshift across a millimetre-scale atomic sample, with fractional frequency uncertainty $7.6\times10^{-21}$ \cite{Bothwell2022}, and across a one-centimetre clock array \cite{Zheng2023}. These results demonstrate extraordinary local potential sensitivity. They do not automatically improve the bound on every dark-energy coupling: a useful scalar test requires two transitions with different sensitivity vectors, a calculable source profile, and control of screening and environmental effects. Space links, including the ACES program, enlarge the potential baseline and provide complementary time-transfer systematics \cite{Savalle2019}.

\subsection{The metric-theory caveat}

In Eq.~\eqref{eq:action} the baryonic sector is metrically coupled, and the only
departure from a metric theory in the visible sector is the gauge kinetic function
$B_F$. Two consequences follow. The achromatic part of the gravitational redshift,
the one that rescales all frequencies by the same factor between two potentials,
is the metric one and is not modified. What $B_F$ adds is a chromatic term: because
$\phi$ takes different values at different potentials, $\alpha$ does too, and a
transition with sensitivity $K_\alpha$ shifts by $K_\alpha\,\delta\ln\alpha$
relative to one with a different $K_\alpha$. Local-position-invariance tests in this
model therefore measure $\zeta_\gamma$ multiplied by the scalar charge of the
source, Eq.~\eqref{eq:kalphamap}, and nothing else; a scalar-tensor theory with
universal $A_b\ne1$ and $B_F=1$ would instead modify the expansion while leaving
all clock ratios unchanged.

A second point concerns screening, which we now define because it is invoked in
Sec.~\ref{sec:taxonomy} as an alternative to the present model. By a screened
model we mean one in which the effective potential
$V_{\rm eff}(\phi;\rho)=V(\phi)+\rho_cA_c(\phi)+\rho_bA_b(\phi)$ has a minimum
$\phi_{\min}(\rho)$ whose position and curvature depend on the ambient density, so
that the field inside a dense body is pinned near $\phi_{\min}$ and its Compton
wavelength there is short. In Eq.~\eqref{eq:action} this does not happen. Baryons do
not source the field beyond $\beta_b^{\rm(em)}$, so the Earth, the Sun, and the
apparatus are transparent to it; and the local dark-matter density
$\rho_c\simeq0.4\,{\rm GeV\,cm^{-3}}$ contributes through Eq.~\eqref{eq:mphi} a mass
whose Compton wavelength is $m_\phi^{-1}\simeq9\,{\rm Mpc}/\beta_c$, far larger than
the Galaxy. The local field therefore follows the cosmological solution, and the
field is a light, unscreened dilaton on all scales from the laboratory to the
Earth--satellite distance relevant for MICROSCOPE. This is the regime in which the
equivalence-principle bounds of Sec.~\ref{sec:clocks}\,D, which are derived for a
massless scalar \cite{DamourDonoghue2010,Berge2018}, apply. Chameleon, symmetron,
and similar mechanisms require a different $V$ or a universal matter coupling and
are outside the model; the corresponding row of Table~\ref{tab:models} states what
would replace Eqs.~\eqref{eq:theta} and~\eqref{eq:kalphamap} in that case
\cite{LevyUzan2025}.

This point is reinforced by weak-equivalence-principle experiments. Electromagnetic and mass couplings generically induce composition-dependent accelerations, and the MICROSCOPE final result constrains the E\"otv\"os parameter at the $10^{-15}$ level \cite{Touboul2022}. Clock drift, clock potential modulation, and differential acceleration probe related but nonidentical charge combinations. Those combinations are related, however, once the electromagnetic charge of each body is assigned explicitly. Section~\ref{sec:clocks}\,D does this and places the three channels on a common footing.

\subsection{Quantitative WEP and source-charge comparison}

Consider first the photon-only case of Eq.~\eqref{eq:action}, which is unscreened by the argument of Sec.~\ref{sec:clocks}\,C. The scalar charge of a neutral body $A$ inherited from its electromagnetic binding energy is
\begin{equation}
 \beta_A^{({\rm em})}=\zeta_\gamma Q_e^A,
 \qquad
 Q_e^A\equiv\frac{\partial\ln m_A}{\partial\ln\alpha},
\label{eq:emcharge}
\end{equation}
where the semi-empirical dilaton charge is approximately \cite{DamourDonoghue2010}
\begin{equation}
 Q_e\simeq10^{-4}\!\left[-1.4+8.2\frac ZA
 +7.7\frac{Z(Z-1)}{A^{4/3}}\right].
\label{eq:Qe}
\end{equation}
For a range longer than the Earth radius,
\begin{equation}
 \eta_{AB}\simeq2\zeta_\gamma^2Q_e^\oplus
 \Delta Q_e^{AB}.
\label{eq:etaem}
\end{equation}
Using the MICROSCOPE alloy values $|\Delta Q_e^{{\rm Pt},{\rm Ti}}|\simeq1.94\times10^{-3}$, $Q_e^\oplus\simeq1.90\times10^{-3}$, and the combined two-sided 95\% endpoint $|\eta|<6.88\times10^{-15}$ gives
\begin{equation}
 |\zeta_\gamma|<3.1\times10^{-5}
 \quad({\rm photon\ only}).
\label{eq:zetawep}
\end{equation}
This full-charge photon-only benchmark is the upper edge of the WEP mapping band in Fig.~\ref{fig:zeta}. A frequently used ``Class I'' varying-$\alpha$ mapping instead uses $\eta\simeq10^{-3}\zeta_\gamma^2$ \cite{Martins2015}. With the same endpoint it gives $|\zeta_\gamma|<2.6\times10^{-6}$ and forms the lower edge of that band. It is stronger because it assumes correlated scalar dependences of nuclear masses; it is not the photon-only limit in Eqs.~\eqref{eq:emcharge}--\eqref{eq:etaem}. This order-of-magnitude spread is precisely why a WEP prior cannot be advertised as a universal limit on $\zeta_\gamma$.

The photon-only number also differs from the single-coupling recast in
Ref.~\cite{Archidiacono2022}. With
$|\zeta_\gamma|=|d_e|/\sqrt{2}$, its $|d_e|<9.9\times10^{-5}$ corresponds to
$|\zeta_\gamma|<7.0\times10^{-5}$, a factor $2.3$ weaker than
Eq.~\eqref{eq:zetawep}. This is not a normalization discrepancy.
Nor is it mainly a charge-model effect. Repeating
Eqs.~\eqref{eq:Qe}--\eqref{eq:zetawep} with the Coulomb-only charge
$Q'_e\simeq7.7\times10^{-4}Z(Z-1)A^{-4/3}$ used in Ref.~\cite{Archidiacono2022}
gives $Q_e^\oplus=1.63\times10^{-3}$ and $|\Delta Q_e^{\rm Pt,Ti}|=2.05\times
10^{-3}$, hence $|\zeta_\gamma|<3.2\times10^{-5}$: the two additional terms of
Eq.~\eqref{eq:Qe} raise the source charge and lower the charge difference, and
largely cancel. The factor comes instead from which form of the MICROSCOPE
result is used. Reference~\cite{Archidiacono2022} works from the compact dilaton
combination introduced for the first MICROSCOPE analysis in Ref.~\cite{Berge2018},
$|d_g^\ast(d_{\hat m}-d_g+0.62\,d_e)|<1.7\times10^{-12}$, where $d_g$ and $d_{\hat m}$ are the dilaton couplings to the gluon and light-quark-mass operators and $d_g^\ast$ is the composition-independent part of the macroscopic charge, with
$d_g^\ast\simeq2.7\times10^{-4}d_e$, which for a photon-only coupling gives
$|d_e|<1.0\times10^{-4}$, whereas Eq.~\eqref{eq:etaem} evaluates $\eta$ directly
against the two-sided final endpoint. The parametrised combination is the weaker
of the two by a factor $2.4$, and that is the whole of the difference.

The same charge assignment closes the loop on annual potential modulation. Solar H/He abundances and Eq.~\eqref{eq:Qe} give $Q_e^\odot\simeq6.5\times10^{-4}$, hence $\beta_s\simeq Q_e^\odot\zeta_\gamma$ in the photon-only limit. Equation~\eqref{eq:kalphamap} then implies
\begin{equation}
 |\zeta_\gamma|<2.5\times10^{-3}
 \quad(k_\alpha,{\rm\ photon\ only}),
\label{eq:zetakalpha}
\end{equation}
far weaker than either Eq.~\eqref{eq:theta95} for a rolling field or Eq.~\eqref{eq:zetawep}. Potential modulation becomes competitive only when the source has a larger non-electromagnetic scalar charge; in that same regime the corresponding baryonic charge must be tested jointly against MICROSCOPE rather than treated as an independent free enhancement.

The qualitative conclusion that atomic-clock drift can constrain a photon-like
dilaton coupling more strongly than equivalence-principle tests was already
obtained in Ref.~\cite{Archidiacono2022}. The purpose of
Eqs.~\eqref{eq:emcharge}--\eqref{eq:zetakalpha} is instead to make the
macroscopic charge assignment explicit and to compare secular drift, annual
potential modulation, and differential free fall within one convention, without
imposing a microscopic relation between the visible and dark couplings.

\subsection{Combining the two clock channels}
\label{sec:joint2d}

\begin{figure}[t]
 \includegraphics[width=\columnwidth]{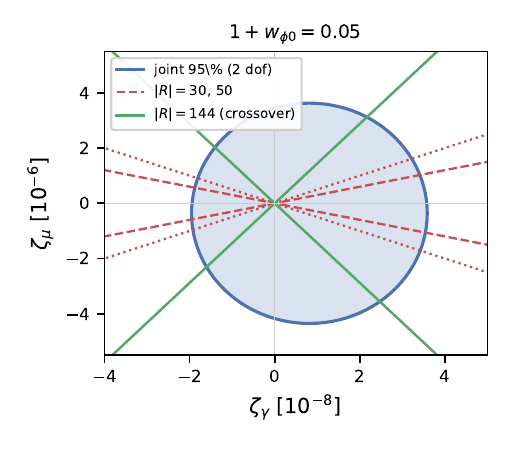}
 \caption{Joint 95\% region for the two visible couplings from the drift results
 of Eqs.~\eqref{eq:clockdrift}, for $1+w_{\phi0}=0.05$ and
 $\Omega_{\phi0}=0.685$. The ellipse is Eq.~\eqref{eq:jointellipse}; being a
 two-parameter region it is wider than the one-dimensional endpoints of
 Eqs.~\eqref{eq:theta95} and~\eqref{eq:thetamu95}. The red rays are the
 unification benchmarks $|R|=30$ and $50$; they leave the region through its
 $\zeta_\gamma$ edges, so the $\alpha$ channel sets the bound. The green rays
 mark the crossover of Eq.~\eqref{eq:Rcross}, beyond which the $\mu$ channel
 would take over.}
 \label{fig:plane}
\end{figure}

Equations~\eqref{eq:theta} and~\eqref{eq:thetamu} constrain two independent
charges with two independent measurements, so the natural object is a region in
the $(\zeta_\gamma,\zeta_\mu)$ plane rather than a pair of separate limits. At
fixed $\varphi'$ the two drift results are uncorrelated Gaussians, with central
values and uncertainties $\sigma_\alpha$, $\sigma_\mu$ as in
Eqs.~\eqref{eq:clockdrift}, and the joint
95\% region ($\Delta\chi^2=5.99$ for two parameters) is the ellipse
\begin{equation}
 \left(\frac{H_0\zeta_\gamma\varphi'-\overline{\dot\alpha/\alpha}}
 {\sigma_\alpha}\right)^{2}
 +\left(\frac{H_0\zeta_\mu\varphi'-\overline{\dot\mu/\mu}}
 {\sigma_\mu}\right)^{2}\le 5.99 .
\label{eq:jointellipse}
\end{equation}
Figure~\ref{fig:plane} shows it for the benchmark $1+w_{\phi0}=0.05$. A
two-parameter region is wider than a one-parameter interval, so the marginal
limits $|\zeta_\gamma|<3.58\times10^{-8}$ and $|\zeta_\mu|<4.35\times10^{-6}$
are weaker than the single-channel values of Sec.~\ref{sec:clocks}\,A by
$\sqrt{5.99}/1.96=1.25$.

A unification relation $\zeta_\mu=R\,\zeta_\gamma$ is a ray through this plane,
and profiling Eq.~\eqref{eq:jointellipse} along it gives the limit a
correlated-coupling model would actually obtain. For $|R|=30$, $40$ and $50$ we
find $|\zeta_\gamma|<3.41$, $3.32$ and $3.22\times10^{-8}$, against
$3.57\times10^{-8}$ as $R\to0$. Imposing the usual unification benchmark
therefore tightens the photon-coupling limit by less than ten percent: in a
joint fit the $\mu$ channel is currently carrying almost no weight, even in a
model that correlates the two charges.

The crossover is set by the ratio of the two measurement uncertainties. The
$\mu$ channel dominates the profiled bound only for
\begin{equation}
 |R|\gtrsim\frac{\sigma_\mu}{\sigma_\alpha}=144 ,
\label{eq:Rcross}
\end{equation}
roughly three times the range spanned by the grand-unified benchmarks. This is
the quantitative form of the statement that $\mu$ is a necessary but not yet
competitive channel: parity requires either a QCD response factor near $10^2$
or a comparable improvement in $\dot\mu/\mu$ sensitivity. Equation~\eqref{eq:Rcross}
is independent of $\varphi'$, of the scalar potential, and of $\Omega_\phi$,
because the field velocity cancels in the ratio; the benchmark
$1+w_{\phi0}$ enters only as an overall dilation of both axes by
$[3\Omega_{\phi0}(1+w_{\phi0})]^{-1/2}$.

\begin{table*}[t]
\caption{Representative clock and redshift observables. Quoted numbers are published measurements or published forecast statements; their dark-energy interpretations require the model assumptions in the last column.}
\label{tab:observables}
\begin{ruledtabular}
\begin{tabular}{p{0.19\textwidth} p{0.27\textwidth} p{0.20\textwidth} p{0.27\textwidth}}
Observable & Representative result & Directly constrains & Dark-energy interpretation \\
\hline
Long-term optical-clock ratios & $\dot\alpha/\alpha=(1.8\pm2.5)\times10^{-19}\,{\rm yr}^{-1}$ \cite{Filzinger2023}; $\dot\mu/\mu=(-8\pm36)\times10^{-18}\,{\rm yr}^{-1}$ \cite{Lange2021} & $\Theta_\alpha=\zeta_\gamma\varphi'$ and $\Theta_\mu=\zeta_\mu\varphi'$ & Coupling--velocity products; a relation between channels is an additional model assumption \\
Annual solar-potential modulation & $k_\alpha=(-2.4\pm3.0)\times10^{-9}$ \cite{Filzinger2023} & Transition response to a local scalar profile & $\zeta_\gamma\beta_s$ only for a light, unscreened, specified source charge \\
Differential free fall & $\eta({\rm Ti,Pt})=(-1.5\pm2.3_{\rm stat}\pm1.5_{\rm syst})\times10^{-15}$ \cite{Touboul2022} & Source charge times a difference of test-body charges & Full-charge photon-only benchmark $|\zeta_\gamma|<3.1\times10^{-5}$; alternative dilaton recasts and correlated matter couplings change the map \\
Satellite clock redshift & $\alpha_{\rm LPI}=(0.19\pm2.48)\times10^{-5}$ \cite{Delva2018} & Departure from modeled metric redshift & Transition- and source-dependent scalar charge or nonmetric coupling \\
Vertical optical-clock comparisons & $7.6\times10^{-21}$ fractional uncertainty within a millimetre-scale sample \cite{Bothwell2022}; centimetre-array gradient consistent with GR \cite{Zheng2023} & Local potential differences and differential transition response & Powerful only after modelling spatial scalar profile and screening \\
Galaxy cross-correlation dipole & BOSS CMASS reported a small-scale asymmetry at $2.7\sigma$ and dipole at $2.8\sigma$ \cite{Alam2017}; an SKA2-like forecast gives up to $\sim50\%$ EFT-parameter improvement \cite{Castello2024JCAP} & Relativistic odd multipoles and tracer Euler equation & Existing small-scale attempt plus a linear-regime dark-force forecast; neither is a present $\beta_c$ measurement \\
\end{tabular}
\end{ruledtabular}
\end{table*}

\section{Cosmological gravitational redshift}
\label{sec:lss}

\subsection{The antisymmetric galaxy dipole}

Throughout, ``dipole'' denotes the $\ell=1$ Legendre multipole of the cross-correlation function $\xi_{AB}(r,\cos\theta)$ of two galaxy populations $A$ and $B$ with respect to the angle $\theta$ between the pair separation and the line of sight, computed in a homogeneous and isotropic background. It exists because $\xi_{AB}\neq\xi_{BA}$ when the two populations have different bias, velocity, or response to the potential, and it vanishes identically in any auto-correlation \cite{Bonvin2014}. It is a property of the two-point statistics of the tracers, not of the spacetime: it neither requires nor indicates any anisotropy of the background, and is unrelated to the Copernican principle.

Observed galaxy redshifts receive Doppler, gravitational-potential, integrated, lensing, and volume-distortion contributions. Standard density and redshift-space-distortion terms dominate the even multipoles of an auto-correlation. Relativistic terms suppressed by $\mathcal{H}/k$ generate an imaginary, odd component in the cross-power spectrum of two populations $A$ and $B$. Schematically,
\begin{equation}
 P_{\rm odd}^{AB}(k,\mu)=i\mu\frac{\mathcal H}{k}
 \left[\mathcal C_A b_B-\mathcal C_B b_A\right]P_m(k)+\cdots,
\label{eq:oddpower}
\end{equation}
where $b_A,b_B$ are tracer biases and $\mathcal C_i$ collects velocity, acceleration, evolution-bias, and magnification-bias contributions \cite{Bonvin2014}. Its configuration-space counterpart begins with a dipole. The signal is often called the gravitational-redshift dipole because the potential difference between the two populations is one of its physical sources, although a gauge-invariant prediction necessarily combines the relevant relativistic terms.

Equation~\eqref{eq:oddpower} uses the standard conformal-time notation,
$\mathcal H=aH$. In Eqs.~\eqref{eq:eulerb} and \eqref{eq:eulerc} below we
switch to cosmic time: overdots denote $\dd/\dd t$, $H=\dot a/a$, and
$\bm\nabla$ denotes the physical spatial gradient, so no explicit factor of
$a^{-1}$ appears in the schematic Euler equations.

The key model-discriminating information is in the Euler equation. Writing $\Psi$ for the Newtonian-gauge time-time metric potential and $\delta\phi$ for the scalar perturbation, baryons minimally coupled to the metric obey, schematically,
\begin{equation}
 \dot{\bm v}_b+H\bm v_b=-\bm\nabla\Psi,
\label{eq:eulerb}
\end{equation}
whereas coupled CDM obeys
\begin{equation}
 \dot{\bm v}_c+(H+\Gamma_c)\bm v_c
 =-\bm\nabla\Psi-\frac{\beta_c}{\mpl}\bm\nabla\delta\phi.
\label{eq:eulerc}
\end{equation}
For Eq.~\eqref{eq:action} the extra drag is $\Gamma_c=\beta_c\dot\phi/\mpl=H\xi_c$ (Appendix~\ref{app:fieldeqs}); it is the momentum-space counterpart of the mass variation in Eq.~\eqref{eq:rhoc}, and takes other forms in the models of Sec.~\ref{sec:taxonomy}. Lensing measures combinations of metric potentials, while tracer velocities respond to the total acceleration. It is important to be precise about what the dipole then constrains. It is sensitive to an effective dark-fifth-force combination built from $\beta_c^2$, the scalar range, the time evolution of the coupling, the halo response and the tracer biases. It does not measure the background transfer rate $\xi_c=\beta_c\varphi'$. The cancellation of $\varphi'$ in Eq.~\eqref{eq:master} therefore requires $\xi_c$ from the background and its perturbations, not from the dipole: clock drift combined with the dipole alone still requires $\varphi'$, as Eq.~\eqref{eq:Hboundbeta} makes explicit through $w_\phi$ and $\Omega_\phi$. A dark fifth force can therefore imitate modified gravity in standard redshift-space distortion and lensing observables, yet fail the relation between metric redshift and motion that enters the dipole.

Equation~\eqref{eq:eulerb} sets the fundamental baryonic coupling to zero, although Eq.~\eqref{eq:emcharge} shows that $\zeta_\gamma\ne0$ generically induces a small composition-dependent $\beta_b$. Galaxies nevertheless move predominantly as CDM-dominated halos, with their luminous baryons tracing the halo bulk velocity. The approximation is safe for the photon-only charge implied by the clock bound, but a larger correlated baryonic coupling can make the tracer acceleration interpolate between Eqs.~\eqref{eq:eulerb} and \eqref{eq:eulerc}, degrading the differential signal; such a model must include its WEP likelihood and a halo-level baryon--CDM response.

\subsection{Forecasted IDE sensitivity}

The observable is no longer purely prospective. A BOSS CMASS analysis reported a cross-correlation asymmetry at $2.7\sigma$ and a small-scale dipole at $2.8\sigma$, approximately consistent with a relativistic halo-model prediction but consistent with zero above $25\,h^{-1}{\rm Mpc}$ \cite{Alam2017}. This is an important measurement attempt, not a linear-regime detection of a dark force. DESI-specific case studies have since developed estimators and nuisance requirements for the relativistic dipole \cite{BonvinDESI2023}.

Within the effective theory of interacting dark energy, an SKA2-like cross-correlation forecast found that adding the dipole could improve constraints on selected EFT parameters by up to about 50\% for two populations whose bias difference is one \cite{Castello2024JCAP}. The improvement arises from degeneracy breaking, not from a dipole signal-to-noise ratio larger than that of the standard multipoles. In a coupled-quintessence versus generalized-Brans--Dicke comparison, synthetic SKA2 data showed that the dipole can distinguish a dark force from modified gravity for sufficiently large coupling and range \cite{Castello2024PRD}.

Reference~\cite{Castello2024PRD} does not forecast a constant $\beta_c$. Its
coupled-quintessence parameter obeys
$\widetilde\beta_2(a)=\beta_{\star2}\sqrt{1-(a_\star/a)^3}$, with $\beta_{\star2}$
an amplitude and $a_\star=0.5$ the scale factor at which the coupling switches
on, and its normalization satisfies
$\widetilde\beta_2=(A_{,\phi}/A)/\sqrt{8\pi G}=\beta_c(a)$ in our conventions.
The reported transition $\beta_{\star2}=0.6$--$0.7$ therefore corresponds at
$a=1$ to $\beta_{c0}=0.561$--$0.655$. Since the likelihood depends on the full
time evolution and on the scalar range, this present-epoch conversion does not
turn the forecast into a constraint on a constant coupling. We display it only
as a contextual toy benchmark for the $10\,{\rm Mpc}$ example.

Existing CMB and BAO analyses constrain a long-range dark fifth force at the sub-percent level in the pure fifth-force realization, $\beta<0.0054$, and at $\beta<0.0287$ in the coupled-dark-energy realization \cite{Archidiacono2022}. Neither number converts directly into our $\beta_c$. The parameter constrained there is $\beta=G_s/(4\pi G_N)$, where $G_s$ is the coupling constant of a dimensionless mediator $s$ whose exchange gives the dark matter a field-dependent mass $m_\chi(s)$. The combination entering the sub-horizon growth equations is $\beta\,(\partial\ln m_\chi/\partial s)^2$, so that
\begin{equation}
 2\beta_c^2=\beta\left(\frac{\partial\ln m_\chi}{\partial s}\right)^{2}.
\label{eq:archimap}
\end{equation}
In the coupled-dark-energy case the requirement that the scalar supply the dark energy forces the background field $\bar s$ to exceed unity and hence $\partial\ln m_\chi/\partial s<1$, a suppression that Ref.~\cite{Archidiacono2022} discusses explicitly and that depends on the initial field value and on the mediator mass. Quoting $|\beta_c|<\sqrt{\beta/2}$ would therefore misstate the constraint. A quantitative overlay on Fig.~\ref{fig:hierarchy} requires the background solution of the specific model, which we do not attempt here.

The cosmological dipole is complementary to local clocks in three ways. First, it can be sensitive when $\zeta_\gamma$ is essentially zero. Second, it probes spatial gradients and the tracer equation of motion rather than only the homogeneous field velocity. Third, it is evaluated in the cosmological environment and is less vulnerable to Solar-System screening, although nonlinear halo screening, galaxy bias, wide-angle terms, and light-cone effects must be included.

\section{Joint constraints}
\label{sec:joint}

\subsection{The visible-to-dark hierarchy}

Combining Eqs.~\eqref{eq:xic} and \eqref{eq:theta} gives the central consistency relation
\begin{equation}
 \boxed{\quad
 \Theta_\alpha=\frac{\zeta_\gamma}{\beta_c}\,\xi_c,
 \qquad
 \Rvd\equiv\left|\frac{\zeta_\gamma}{\beta_c}\right|
 =\left|\frac{\Theta_\alpha}{\xi_c}\right| .\quad}
\label{eq:master}
\end{equation}
We call this the hierarchy relation. It is an identity within Eq.~\eqref{eq:action}, whether or not $\alpha$ varies: the baryonic transfer induced by $\beta_b^{\rm(em)}$, $Q_b=\zeta_\gamma Q_e^{b}\rho_b\dot\phi/\mpl$, does not enter it, and its ratio to $Q_c$ is $(\zeta_\gamma Q_e^b/\beta_c)(\rho_b/\rho_c)\lesssim10^{-10}$ for the bounds obtained below. The relation is independent of the scalar potential, but not model independent. Six conditions must hold before it is applied to a cosmological posterior. Each
is stated as a question that has to be answered affirmatively:
\begin{enumerate}
\setlength{\itemsep}{2pt}
 \item Is the field that changes $\alpha$ identical to the mediator that
 generates $Q_c$?
 \item Is $\alpha(\phi)$ linear over the present clock baseline, and is
 $A_c(\phi)$ well approximated by a constant $\beta_c$?
 \item Does the local field track the homogeneous cosmological solution, rather
 than a screened or virialized profile?
 \item Is the fitted $Q_c$ a conformal mass-variation term proportional to
 $\rho_c\dot\phi$, rather than momentum exchange or a fluid-level ansatz?
 \item Are the scalar normalization, the sign of $Q$, and the matter-frame
 definitions consistent between the clock and cosmological calculations?
 \item Does the redshift dependence match? In particular, has a posterior for
 constant phenomenological $\xi_{\rm DE}$ been refitted with the evolution
 $\xi_c(a)=\beta_c\varphi'(a)$ implied by constant $\beta_c$, rather than
 identified only at $z=0$?
\end{enumerate}
When all six hold, a cosmological determination of $\xi_c$ and a clock null
result bound the hierarchy directly, without separately measuring $w$ or
$\varphi'$. A negative answer does not make clocks irrelevant; it means that the
relevant local solution or operator-specific response must replace the one-line
mapping.

For example, Eq.~\eqref{eq:theta95} implies
\begin{equation}
 \Rvd<1.94\times10^{-7}
 \left(\frac{|\xi_c|}{0.05}\right)^{-1}
\label{eq:Hboundxic}
\end{equation}
at the two-sided 95\% endpoint. Thus, a five-percent transfer per Hubble time mediated by the same scalar would require the photon coupling to be suppressed by nearly seven orders of magnitude relative to the CDM coupling. A symmetry, sequestering mechanism, or environment-dependent suppression would then be part of the model rather than an optional detail.

\subsection{Canonical quintessence mapping}

\begin{figure}[t]
 \includegraphics[width=\columnwidth]{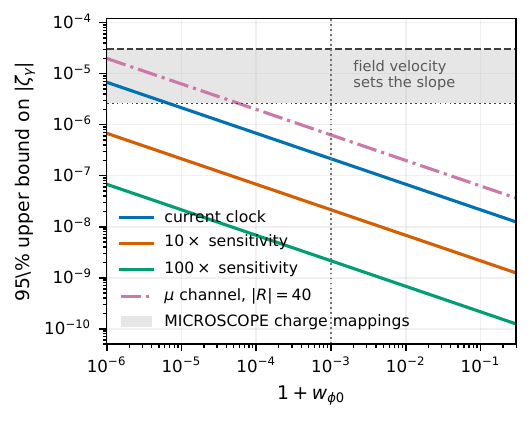}
 \caption{Conditional 95\% upper bounds on the electromagnetic coupling of a canonical scalar for $\Omega_{\phi0}=0.685$. The solid blue curve uses the $\alpha$ endpoint in Eq.~\eqref{eq:theta95}; projected curves divide that endpoint by 10 or 100, exactly as in Table~\ref{tab:numbers}. The dash-dotted curve uses the $\mu$ endpoint with the benchmark $|R|=40$. The gray band spans the two WEP mappings discussed in Sec.~\ref{sec:clocks}\,D: the correlated ``Class I'' value $|\zeta_\gamma|<2.6\times10^{-6}$ at its lower edge and the full-charge photon-only value $|\zeta_\gamma|<3.1\times10^{-5}$ at its upper edge. Its width represents charge-model dependence, not a statistical confidence interval. These are conditional sensitivity comparisons, not new likelihoods.}
 \label{fig:zeta}
\end{figure}

For a canonical scalar,
\begin{equation}
 \varphi'^2=3\Omega_\phi(1+w_\phi),
\label{eq:phiprime}
\end{equation}
which requires $1+w_\phi\ge0$. The DESI DR2 $w_0w_a$ posterior favors $w_0>-1$ and $w_a<0$, so parts of its preferred history cross into $w<-1$ and cannot be represented by Eq.~\eqref{eq:phiprime} for one canonical scalar \cite{DESI2025DR2}. The values of $1+w_{\phi0}$ used below are therefore transparent rolling-field benchmarks, not samples drawn from the DESI $w_0w_a$ posterior. Under this restriction, the clock limit becomes
\begin{align}
 |\zeta_\gamma|_{95}
 <{}&\frac{9.72\times10^{-9}}
 {\sqrt{3\Omega_{\phi0}(1+w_{\phi0})}}\nonumber\\
 ={}&3.03\times10^{-8}
 \left(\frac{1+w_{\phi0}}{0.05}\right)^{-1/2}
 \left(\frac{\Omega_{\phi0}}{0.685}\right)^{-1/2}.
\label{eq:zetabound}
\end{align}
Figure~\ref{fig:zeta} shows this conditional limit. The divergence as $w\rightarrow-1$ is physical: a frozen scalar cannot reveal its coupling through a secular drift. Conversely, a prior on $\zeta_\gamma$ is indispensable if one wishes to turn a clock into a bound on $w+1$. This is why very strong projected clock-only limits on the dark-energy equation of state can depend more on the assumed visible coupling than on timing sensitivity \cite{Elder2025,Trivedi2025}.

\begin{figure}[t]
 \includegraphics[width=\columnwidth]{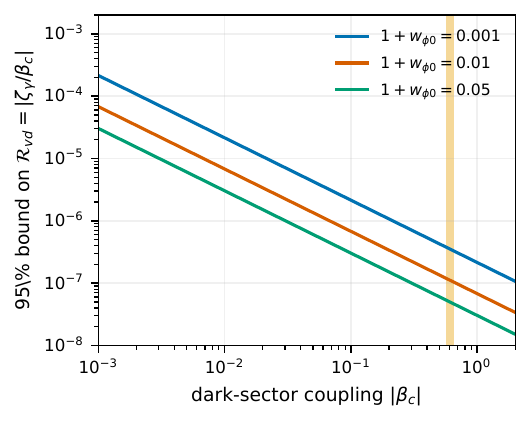}
 \caption{Clock-implied upper bound on the visible-to-dark hierarchy as a function of $|\beta_c|$ for canonical scalar trajectories. The orange band maps the $\beta_{\star2}=0.6$--$0.7$ discrimination transition in the $\lambda_*=10\,{\rm Mpc}$ SKA2 synthetic-data example of Ref.~\cite{Castello2024PRD} to its present-epoch value, $\beta_{c0}=0.561$--$0.655$. The overlay is specific to that time-dependent model and scalar range and is not a general constraint.}
 \label{fig:hierarchy}
\end{figure}

If a gravitational-redshift or other cosmological analysis measures $\beta_c$, Eqs.~\eqref{eq:phiprime} and \eqref{eq:zetabound} give
\begin{equation}
 \Rvd<\frac{9.72\times10^{-9}}
 {|\beta_c|\sqrt{3\Omega_{\phi0}(1+w_{\phi0})}}.
\label{eq:Hboundbeta}
\end{equation}
The dependence is shown in Fig.~\ref{fig:hierarchy}. For
$1+w_{\phi0}=0.05$, evaluating Eq.~\eqref{eq:Hboundbeta} at the upper edge
$\beta_{\star2}=0.7$, corresponding to $\beta_{c0}=0.655$, gives
$\Rvd<4.63\times10^{-8}$. Across the mapped $0.561$--$0.655$ band the
corresponding limits range from $5.40\times10^{-8}$ to $4.63\times10^{-8}$.
This remains an algebraic illustration, since the time-dependent model of
Ref.~\cite{Castello2024PRD} is not the constant-coupling canonical trajectory
assumed in Eq.~\eqref{eq:phiprime}. For comparison, a constant
$|\beta_c|=0.05$ gives $\Rvd<6.06\times10^{-7}$.

\begin{figure}[t]
 \includegraphics[width=\columnwidth]{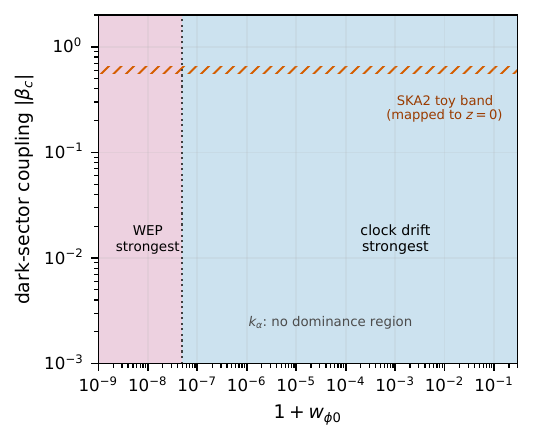}
 \caption{Probe-dominance and coverage map in the unscreened, long-range limit. Background color compares the two WEP charge mappings in Fig.~\ref{fig:zeta}: WEP is strongest for both below the photon-only crossover, $1+w_{\phi0}\simeq4.9\times10^{-8}$; the winner is charge-map dependent up to the ``Class I'' crossover, $1+w_{\phi0}\simeq6.8\times10^{-6}$; and clock drift is strongest for both above it. The $k_\alpha$ channel is never strongest because Eq.~\eqref{eq:zetakalpha} is weaker. The orange SKA2 strip is the present-epoch mapping $\beta_{c0}=0.561$--$0.655$ of the time-dependent toy discrimination region in Ref.~\cite{Castello2024PRD}. A literal four-way ``winner'' would be ill-defined because the dipole constrains $\beta_c$ while the other probes constrain $\zeta_\gamma$; the overlay displays this orthogonality rather than comparing unlike charges. The MICROSCOPE mapping assumes a scalar range exceeding the orbital radius, $k_\alpha$ a range exceeding $1\,$AU, and the dipole overlay $10\,{\rm Mpc}$.}
 \label{fig:map}
\end{figure}

\subsection{Mapping a phenomenological interaction}

If, and only if, Eq.~\eqref{eq:Qphen} is completed by the same conformal scalar used in the clock sector, Eqs.~\eqref{eq:master} and \eqref{eq:ximap} yield the present-epoch identity
\begin{equation}
 \Rvd<9.72\times10^{-9}
 \left|\xi_{\rm DE}\frac{\Omega_{\rm DE}}{\Omega_c}\right|^{-1}.
\label{eq:HboundxiDE}
\end{equation}
For $\Omega_{\rm DE}=0.685$ and $\Omega_c=0.265$, $|\xi_{\rm DE}|=0.022$, $0.05$, and $0.132$ correspond to $\Rvd<1.71\times10^{-7}$, $7.52\times10^{-8}$, and $2.85\times10^{-8}$, respectively. These values illustrate the coupling scales encountered in recent DESI DR2 IDE fits \cite{Silva2025}. They do not reproduce or combine the cosmological posteriors: the fitted $\xi_{\rm DE}$ is constant across redshift and its posterior is driven by the full expansion history, whereas Eq.~\eqref{eq:HboundxiDE} evaluates an instantaneous $z=0$ relation. It must not be read as a bound on a constant $\beta_c$ without refitting the redshift evolution implied by $\beta_c\varphi'(a)$.

\begin{figure}[t]
 \includegraphics[width=\columnwidth]{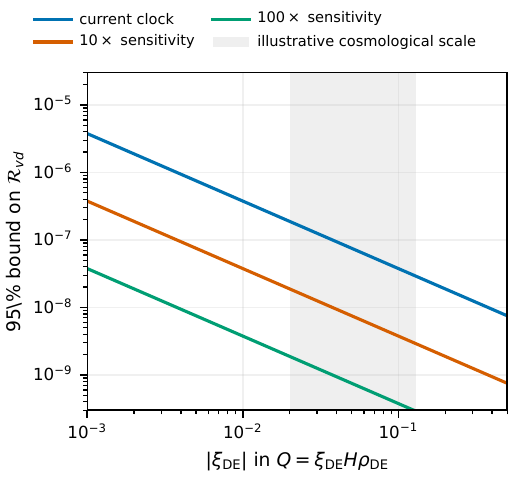}
 \caption{Present-epoch conditional mapping for $Q_c=\xi_{\rm DE}H\rho_{\rm DE}$, assuming its microphysical completion is precisely Eq.~\eqref{eq:conservation}. The gray region marks $0.02<|\xi_{\rm DE}|<0.13$ only as an illustrative cosmological scale motivated by recent fits; a constant-$\xi_{\rm DE}$ posterior cannot be converted into constant $\beta_c$ by this plot alone.}
 \label{fig:xi}
\end{figure}

The cancellation in Eq.~\eqref{eq:master} happens for a simple reason: both observables are one charge multiplied by the same field velocity, so their ratio is the ratio of the charges and the trajectory drops out. That is useful, but it is also diagnostic. A model with the phenomenological background transfer of Eq.~\eqref{eq:Qphen} and a frozen canonical field would require $\beta_c$ to grow as $1/\varphi'$ to keep $Q_c$ fixed. That regime can invalidate the constant-coupling expansion and generate a large fifth force. The algebraic mapping should therefore be checked against perturbative stability, structure growth, and local screening, not imposed as a post-processing identity on an arbitrary IDE chain.

\begin{table}[t]
\caption{Selected conditional 95\% limits from the clock endpoints. ``$10\times$'' divides the corresponding current endpoint by ten. The $\mu$ row additionally assumes $\zeta_\mu=R\zeta_\gamma$ with $|R|=40$.}
\label{tab:numbers}
\begin{ruledtabular}
\begin{tabular}{p{0.48\columnwidth}cc}
Assumption & Current & $10\times$ \\
\hline
$1+w_{\phi0}=10^{-3}$: $|\zeta_\gamma|$ & $2.14\times10^{-7}$ & $2.14\times10^{-8}$ \\
$1+w_{\phi0}=10^{-2}$: $|\zeta_\gamma|$ & $6.78\times10^{-8}$ & $6.78\times10^{-9}$ \\
$1+w_{\phi0}=0.05$: $|\zeta_\gamma|$ & $3.03\times10^{-8}$ & $3.03\times10^{-9}$ \\
$1+w_{\phi0}=0.05$, $\mu$ channel, $|R|=40$: $|\zeta_\gamma|$ & $8.89\times10^{-8}$ & $8.89\times10^{-9}$ \\
$|\beta_c|=0.05$, $1+w_{\phi0}=0.05$: $\Rvd$ & $6.06\times10^{-7}$ & $6.06\times10^{-8}$ \\
$|\xi_c|=0.05$: $\Rvd$ & $1.94\times10^{-7}$ & $1.94\times10^{-8}$ \\
$|\xi_{\rm DE}|=0.05$: $\Rvd$ & $7.52\times10^{-8}$ & $7.52\times10^{-9}$ \\
\end{tabular}
\end{ruledtabular}
\end{table}

\section{Which dark-energy models are tested?}
\label{sec:taxonomy}

Everything in Secs.~\ref{sec:framework}--\ref{sec:joint} holds for the model of Eq.~\eqref{eq:action}. Table~\ref{tab:models} lists model classes that do \emph{not} reduce to it and states, for each, which of the assumptions used above fails and what would have to replace it; it is a map of the boundary of the analysis, not an extension of its results. The decisive question is not merely whether a model contains a scalar, but whether the same scalar controls the relevant dimensionless constant, couples to the matter sourcing the local profile, and acts on the cosmological tracers.

\begin{table*}[t]
\caption{Applicability of clock and gravitational-redshift tests to model classes, with the principal failure mode of each conversion. ``Conditional'' means that a mediator and coupling map must be supplied; ``no direct map'' means that the named model class alone makes no clock prediction. All quantitative results of this paper are derived for the model of Eq.~\eqref{eq:action}; this table states, for models outside it, which assumption fails.}
\label{tab:models}
\begin{ruledtabular}
\begin{tabular}{p{0.20\textwidth} p{0.22\textwidth} p{0.22\textwidth} p{0.29\textwidth}}
Model class & Clock drift / local redshift & Cosmological dipole & Main implication or loophole \\
\hline
Cosmological constant & No field-induced signal & Standard relativistic dipole & Null benchmark; clocks do not measure vacuum energy directly \\
Minimally coupled quintessence & No direct clock signal & Standard dipole apart from altered expansion/growth & A dynamical $w$ alone does not imply varying constants \\
Electromagnetically coupled quintessence or dilaton & Yes: $\zeta_\gamma\varphi'$, $\zeta_\mu\varphi'$, and source-dependent LPI & Usually standard unless matter is also coupled & Clocks constrain coupling--velocity products; WEP data constrain charge combinations fixed only after a visible-sector charge assignment \\
CDM-coupled quintessence & Only if the same field also couples to visible parameters & Yes: modified CDM/tracer Euler equation & Clock drift plus an independently inferred $\xi_c$ constrain $\Rvd$; the dipole supplies complementary force information \\
Phenomenological $Q=\xi H\rho_i$ & No direct map & Conditional on perturbation prescription & $\xi$ is not a microscopic charge; a constant-$\xi$ history is not a $z=0$ charge and requires a scalar completion and redshift refit \\
Momentum-exchange or elastic IDE & Usually no homogeneous clock relation & Potentially strong velocity/dipole effects & Background $Q$ may vanish, so the clock--transfer relation does not apply \\
Kinetic or disformal coupling & Possible, with derivative-dependent sensitivities & Scale- and velocity-dependent & Replace constant $\zeta_\gamma,\beta_c$ by the appropriate operators \cite{BarrosFonseca2023,Gleyzes2015} \\
Screened scalar-tensor dark energy & Local drift/profile may be suppressed or enhanced near boundaries & Cosmological effect can remain & Solve the environmental field profile; do not insert the cosmological $\phi(t)$ into a laboratory formula \cite{LevyUzan2025} \\
Early dark energy with varying $\alpha$ & Present clocks constrain only a surviving rolling field/coupling & Model dependent & A phase transition, decay, coupling zero, or tuned trajectory can separate recombination from today \cite{VacherSchoeneberg2024} \\
Multi-field dark sector & Possible but not tied to the dark-force mediator & Possible through a different field & The single-field hierarchy loses meaning; use a coupling matrix and field-space trajectory \\
\end{tabular}
\end{ruledtabular}
\end{table*}

\subsection{Minimally coupled and electromagnetic models}

Neither a cosmological constant nor minimally coupled quintessence produces a clock drift by itself. The scalar may roll while all dimensionless Standard-Model ratios remain fixed. Conversely, a clock signal can arise from a subdominant light scalar that is not dark energy. Establishing a dark-energy origin requires correlation with the cosmological trajectory, redshift evolution, or another observable that measures the field's stress-energy.

For an electromagnetic coupling, the integrated variation is
\begin{equation}
 \frac{\Delta\alpha}{\alpha}(z)
 =\zeta_\gamma\int_0^z
 \sqrt{3\Omega_\phi(z')[1+w_\phi(z')]}
 \frac{\dd z'}{1+z'},
\label{eq:integratedalpha}
\end{equation}
up to the sign of the roll, for a canonical monotonic scalar and a linear gauge kinetic function \cite{Martins2015}. Astrophysical absorption spectra and the CMB extend the time baseline, while clocks anchor the derivative at $z=0$. In unfine-tuned single-field models, the local clock bound can make percent-level changes in $\alpha$ at recombination difficult; a recent analysis obtained the generic benchmark $|\Delta\alpha/\alpha|<5\times10^{-4}$ at recombination under its stated assumptions \cite{VacherSchoeneberg2024}. Kinetic electromagnetic couplings or a nonmonotonic trajectory alter Eq.~\eqref{eq:integratedalpha} \cite{BarrosFonseca2023}.

\subsection{Interacting and screened models}

Runaway dilaton models provide a concrete example in which cosmology, clocks, and equivalence-principle tests must be evolved together \cite{Vacher2023}. Coupled quintessence provides the analogous dark-sector example. If visible couplings are technically natural only at the same order as $\beta_c$, the bound on $\Rvd$ can eliminate a cosmologically interesting interaction. If a symmetry suppresses $\zeta_\gamma$ while leaving $\beta_c$ finite, a null clock result is expected and large-scale gravitational redshift becomes the more direct probe.

Screening breaks the simplest local--cosmological identification. Chameleon, symmetron, dilaton, or Vainshtein-like mechanisms can make the local field value and gradient depend on ambient density, object compactness, vacuum-chamber geometry, or boundary conditions. A clock may then probe a thin-shell profile rather than $\varphi'_0$. Recent work has shown explicitly how gravitational-redshift clock experiments can access regions of screened-scalar parameter space not covered by ordinary fifth-force searches \cite{LevyUzan2025}. Such an experiment remains relevant to dark energy, but Eq.~\eqref{eq:master} must be replaced by a matched cosmological and local solution.

\section{A joint inference program}
\label{sec:program}

A robust analysis should operate at the level of observables and propagate a common microphysical parameter set. One possible parameter vector is
\begin{equation}
 \bm\theta=\{\bm\theta_{\rm cosmo},V(\phi),\zeta_\gamma,
 \zeta_\mu,\beta_b,\beta_c,m_\phi,\bm\theta_{\rm scr}\},
\label{eq:paramvector}
\end{equation}
where $\bm\theta_{\rm scr}$ describes any screening sector. The likelihood can then be factorized conditionally as
\begin{align}
 \mathcal L(\bm\theta)={}&
 \mathcal L_{\rm clock\,drift}
 \mathcal L_{\rm local\,LPI}
 \mathcal L_{\rm WEP}\nonumber\\
 &\times\mathcal L_{\rm even\,LSS}
 \mathcal L_{\rm dipole}
 \mathcal L_{\rm background}.
\label{eq:likelihood}
\end{align}
Equation~\eqref{eq:likelihood} is schematic and its factorization assumes conditional independence. That assumption is nontrivial for the dipole and even multipoles extracted from the same survey volume, which share cosmic variance, selection functions, tracer biases, and calibration systematics; their joint covariance must replace a naive product. Analogous correlations arise among clock ratios that reuse a common reference. The purpose of the equation is to prevent double counting and keep the laboratory response coefficients explicit, not to assert independence by notation.

Several practical requirements follow.

\begin{enumerate}
 \item \emph{Use multiple clock species.} A claimed scalar signal should fit a common set of $\dot\alpha$, $\dot\mu$, and nuclear-parameter drifts through independently calculated $K$ coefficients. Section~\ref{sec:joint2d} quantifies this: $|R|=30$--$50$ improves the $\zeta_\gamma$ limit by under ten percent, and parity requires $|R|\gtrsim144$ or an equivalent gain in $\dot\mu/\mu$ sensitivity.
 \item \emph{Separate secular and potential templates.} A cosmological drift, an annual solar-potential modulation, ultralight-dark-matter oscillations, and environmental systematics have different phases and spectra.
 \item \emph{Model the tracer connection.} A CDM fifth force affects halo formation and velocities; galaxy bias, evolution bias, and magnification bias cannot be treated as external constants in a precision dipole analysis.
 \item \emph{Retain the scalar range.} The background transfer, local profile, and survey-scale force can sample three different effective regimes of $m_\phi$.
 \item \emph{Declare frame and sign conventions.} Bounds on $\beta_c$ and $\xi$ cannot be compared reliably without the scalar normalization and the definition of $Q$.
\end{enumerate}

The most informative outcomes are consistency patterns rather than single upper limits. A cosmological dipole indicating a dark force, accompanied by null clock drift and null LPI modulation, measures or bounds visible-sector sequestering. A clock drift with no dark-force signature favors a visible-coupled scalar, although a non-dark-energy ultralight field must be excluded. Potential modulation without secular drift points toward an environment-sourced profile or an oscillatory field. A discrepancy among clock pairs that does not follow their sensitivity coefficients is a diagnostic of systematics rather than scalar cosmology.

\subsection{Forecast scaling}

Figures~\ref{fig:zeta} and \ref{fig:xi} include simple factors-of-10 and 100 sensitivity improvements. These curves divide the present two-sided endpoint by the stated factor, keeping unchanged cosmological parameters and no theory floor; they are not dates or guaranteed performance. Clock networks such as QSNET are designed to combine atomic, molecular, and highly charged ion responses over a broad range of timescales \cite{Barontini2022}. Oscillatory searches must be fitted separately from secular templates \cite{Filzinger2023,Sherrill2023}. Most concretely, the 2024 laser excitation of the $^{229}{\rm Th}$ nucleus and its direct frequency ratio to an $^{87}{\rm Sr}$ clock have moved enhanced nuclear sensitivity from a proposal to an experimental platform \cite{Tiedau2024,Zhang2024}; converting that sensitivity into $\zeta_\mu$ or QCD-sector charges still requires a validated nuclear response calculation.

For the galaxy dipole, increased volume is not the only lever. Signal strength depends on tracer-bias contrast, number density, redshift coverage, magnification and evolution biases, separation cuts, and control of wide-angle contamination. Joint optimization with even multipoles and lensing is therefore more meaningful than quoting an isolated dipole precision. In particular, the gravitational-redshift observable is valuable because of the direction in parameter space it constrains, even when its raw signal-to-noise is smaller than that of the quadrupole \cite{Castello2024JCAP}.

\section{Discussion}
\label{sec:discussion}

The numerical results expose a severe hierarchy problem for the simplest single-field realization. If the field rolls fast enough to have $1+w_{\phi0}\sim0.01$--$0.1$ and couples to CDM strongly enough to modify cosmic growth measurably, the absence of $\alpha$ drift requires $|\zeta_\gamma|\ll|\beta_c|$. The hierarchy is typically between $10^{-8}$ and $10^{-6}$ for the examples in Table~\ref{tab:numbers}. This conclusion is more robust than a clock-only bound on $w$ because the field velocity cancels when an energy-transfer rate $\xi_c$ is independently determined.

There are several logically distinct escape routes.

\begin{itemize}
 \item \emph{Visible-sector sequestering:} a symmetry or fixed point suppresses $\zeta_\gamma$ and baryonic charges while allowing a dark coupling.
 \item \emph{Cosmological freezing:} $\varphi'_0$ is extremely small. This weakens drift constraints but also suppresses conformal energy transfer unless $\beta_c$ is large or time dependent.
 \item \emph{Environmental screening:} the laboratory field does not follow the cosmological solution. Local redshift tests then constrain the screening profile rather than Eq.~\eqref{eq:theta}.
 \item \emph{Finite range:} the mediator affects one of laboratory, galactic, or cosmological scales but not the others.
 \item \emph{Multiple fields:} one field changes constants and another mediates the dark interaction; $\Rvd$ is replaced by a projection in field space.
 \item \emph{Nonconformal interaction:} momentum exchange, kinetic mixing, or disformal operators invalidate $Q_c=\beta_c\rho_c\dot\phi/\mpl$.
\end{itemize}

These are physical alternatives, not nuisances to be marginalized blindly. The observational program is valuable precisely because it distinguishes them. Local clock drift measures the time direction of the scalar trajectory; potential modulation measures a sourced spatial direction; the cosmological dipole measures differential acceleration on large scales. A viable theory must place all three in a common space-time solution.

The analysis also clarifies what should not be claimed. Clock data do not directly bound a generic $Q=\xi H\rho_i$ model, a metastable vacuum transfer, or elastic dark-sector scattering. Local agreement with the metric redshift does not exclude minimally coupled quintessence. Conversely, a cosmological gravitational-redshift signal is not by itself evidence that constants vary. Translating any of these statements requires a specified mediator and an explicit assignment of its couplings.

\section{Conclusions}
\label{sec:conclusions}

Atomic clocks and gravitational redshift can form a stringent cross-scale test of dark-energy microphysics, provided two meanings of ``gravitational redshift'' and two levels of model description are kept separate.

Long-term clock ratios constrain $\Theta_\alpha=\zeta_\gamma\varphi'$, while local potential modulation constrains a source-dependent combination such as $\zeta_\gamma\beta_s$. The odd dipole of galaxy cross-correlations probes the tracer Euler equation and can identify a dark fifth force that is degenerate with modified gravity in even clustering and lensing observables. In a single conformally coupled scalar completion, clock drift and an independently inferred background transfer obey the relation $\Theta_\alpha=(\zeta_\gamma/\beta_c)\xi_c$. The dipole supplies complementary information on the force response, but does not by itself provide the $\xi_c$ needed for the cancellation of $\varphi'$.

The clock-only quintessence scaling, its equivalence-principle consequences, and
its translation into dark-force bounds in explicit microscopic models have
precedents, notably Ref.~\cite{Archidiacono2022}. The contribution here is the
observable-level separation of $\zeta_\gamma\varphi'$, $\beta_c\varphi'$, and
the force combination beginning at order $\beta_c^2$; the explicit visible
charge assignment for the WEP and potential-modulation limits; the joint
two-coupling region of Sec.~\ref{sec:joint2d}, in which the $\mu$ channel
tightens the photon-coupling limit by under ten percent for the usual
unification benchmarks and reaches parity only for $|R|\gtrsim144$; and the six
conditions under which the instantaneous hierarchy identity can be applied to a
cosmological posterior. In particular, neither the dipole nor a phenomenological
$\xi$ is by itself a measurement of the microscopic background transfer needed
for the trajectory cancellation.

Using the current Yb$^+$ result, we find $|\Theta_{\alpha0}|<9.72\times10^{-9}$ at the two-sided 95\% endpoint. This yields a conditional canonical-quintessence bound $|\zeta_\gamma|<3.03\times10^{-8}$ for $1+w_{\phi0}=0.05$, and a potential-independent hierarchy bound $\Rvd<1.94\times10^{-7}$ if $|\xi_c|=0.05$. A cosmologically relevant dark interaction can therefore coexist with null clock drift in the minimal model only through a highly suppressed visible coupling, a nearly frozen trajectory together with a correspondingly adjusted dark coupling, or a failure of the local--cosmological mapping through screening, range, multiple fields, or nonconformal dynamics.

The appropriate next step is a joint likelihood that evolves a specified scalar theory through background cosmology, perturbations, halo/tracer dynamics, and the local experimental environment. Such an analysis would turn the present hierarchy estimates into model posteriors. Until then, the strongest defensible conclusion is a consistency condition: precision timekeeping does not measure dark energy directly, but it can decide whether a scalar responsible for cosmological dark-sector dynamics is allowed to communicate with the visible world.

\begin{acknowledgments}
The author used an AI assistant (Claude, Anthropic) to check the numerical
results, verify internal consistency, and assist with the editing of the
manuscript. All scientific content, derivations, and conclusions are the
author's own, and the author takes full responsibility for the text.
\end{acknowledgments}

\appendix

\section{Conventions and frame dependence}
\label{app:conventions}

Our operational definition $\dot\rho_c+3H\rho_c=Q_c$ fixes the sign of $Q_c$. Some of the IDE literature uses the opposite sign or assigns $Q$ to the dark-energy continuity equation. All hierarchy limits use absolute values and are unaffected, but a comparison of preferred positive or negative energy flow requires translating conventions.

The scalar normalization is fixed by a canonical kinetic term and the reduced Planck mass in Eq.~\eqref{eq:action}. A coupling denoted $\beta$ elsewhere may refer to $\sqrt{2}\beta_c$, to the dimensionful $\partial\ln A/\partial\phi=\beta_c/\mpl=\beta_c\sqrt{8\pi G}$, or directly to the force fraction $2\beta_c^2$. Numerical comparisons must use Eq.~\eqref{eq:fifthforce} or the complete perturbation equations rather than a symbol match.

Conformal transformations redistribute variation between the metric and matter parameters, but an observable frequency ratio is frame invariant when units and matter responses are transformed consistently. We use the Einstein frame because the distinction between $A_b$ and $A_c$ is explicit. A universal Jordan-frame coupling can preserve LPI even though the corresponding Einstein-frame masses depend on $\phi$; only dimensionless, differential observables should be compared with data.

\section{Numerical prescription}
\label{app:numerics}

For the Gaussian clock result $x=(1.8\pm2.5)\times10^{-19}\,{\rm yr}^{-1}$ \cite{Filzinger2023}, the largest absolute endpoint of its two-sided 95\% interval is
\begin{equation}
 |x|_{95}\equiv|\bar x|+1.96\sigma_x
 =6.70\times10^{-19}\,{\rm yr}^{-1}.
\end{equation}
With
\begin{align}
 H_0&=67.4\frac{{\rm km}}{{\rm s\,Mpc}}
 =2.1843\times10^{-18}\,{\rm s}^{-1}\nonumber\\
 &=6.8931\times10^{-11}\,{\rm yr}^{-1},
\end{align}
one obtains Eq.~\eqref{eq:theta95}. It is simply the upper endpoint for $\bar x>0$, not a bespoke confidence construction. For the $\mu$ result, $|-8|+1.96(36)=78.56$ in units of $10^{-18}\,{\rm yr}^{-1}$ gives Eq.~\eqref{eq:thetamu95}. All curves use $\Omega_{\phi0}=0.685$ and, when Eq.~\eqref{eq:ximap} is invoked, $\Omega_{c0}=0.265$. The plotted improvement factors divide the relevant current endpoint by 10 or 100. Both endpoints are summaries rather than substitutes for a joint likelihood built from the original clock-ratio time series and its correlated systematics.

\section{Field equations of the model}
\label{app:fieldeqs}

This appendix collects the equations that follow from Eq.~\eqref{eq:action} so that
every assumption used in the text can be checked against them. We take
$A_c=\exp(\beta_c\phi/\mpl)$ with constant $\beta_c$, $A_b=1$, and the photon-only
visible sector $\alpha=\alpha_0/B_F(\phi)$; the induced baryonic charge
$\beta_b^{\rm(em)}$ is written explicitly where it enters and then dropped.

\emph{Background.} In a flat FLRW metric $\dd s^2=-\dd t^2+a^2\dd\bm x^2$, with
$H=\dot a/a$ and $\rho_\phi=\tfrac12\dot\phi^2+V$, $p_\phi=\tfrac12\dot\phi^2-V$,
\begin{align}
 3\mpl^2H^2 &=\rho_\phi+\rho_c+\rho_b+\rho_r,\\
 \ddot\phi+3H\dot\phi+V_{,\phi}
 &=-\frac{\beta_c\rho_c+\beta_b^{\rm(em)}\rho_b}{\mpl},\\
 \dot\rho_c+3H\rho_c &=\beta_c\rho_c\frac{\dot\phi}{\mpl}\equiv Q_c,\\
 \dot\rho_b+3H\rho_b &=\beta_b^{\rm(em)}\rho_b\frac{\dot\phi}{\mpl}\equiv Q_b,\\
 \dot\rho_r+4H\rho_r &=0,\\
 \dot\rho_\phi+3H(\rho_\phi+p_\phi) &=-Q_c-Q_b .
\end{align}
The last equation is the scalar equation multiplied by $\dot\phi$; adding all
continuity equations gives $\dot\rho_{\rm tot}+3H(\rho_{\rm tot}+p_{\rm tot})=0$.
Radiation is unaffected because its stress tensor is traceless. The CDM equation
integrates to Eq.~\eqref{eq:rhoc}, and $\rho_c/\rho_b\propto A_c(\phi)$ is the only
background quantity that distinguishes the model from an uncoupled one with the
same $H(a)$. The dimensionless roll is $\varphi'=\dot\phi/(H\mpl)$ and, for a
canonical field, $\varphi'^2=3\Omega_\phi(1+w_\phi)$ with
$\Omega_\phi=\rho_\phi/(3\mpl^2H^2)$ and $1+w_\phi=\dot\phi^2/\rho_\phi$.

\emph{Cosmological redshift.} Photons propagate on $g_{\mu\nu}$ and their momentum
redshifts as $a^{-1}$; for a comoving emitter $1+z=a_0/a$. An atomic transition of
laboratory frequency $\nu_{\rm lab}$ and sensitivity $K_\alpha=\partial\ln\nu/\partial\ln\alpha$
is emitted at frequency $\nu_{\rm lab}[1+K_\alpha\,\delta\ln\alpha(z)]$ with
$\delta\ln\alpha(z)=\ln[B_F(\phi_0)/B_F(\phi(z))]$, so its observed frequency is
$\nu_{\rm lab}(1+z)^{-1}[1+K_\alpha\,\delta\ln\alpha(z)]$. The redshift inferred from
a line with $K_\alpha=0$ is the metric one.

\emph{Perturbations.} In Newtonian gauge,
$\dd s^2=-(1+2\Psi)\dd t^2+a^2(1-2\Phi)\dd\bm x^2$, with CDM density contrast
$\delta_c$, peculiar velocity $\bm v_c$, and scalar perturbation $\delta\phi$, the
CDM fluid equations are
\begin{align}
 \dot\delta_c+\frac{1}{a}\bm\nabla\!\cdot\!\bm v_c
 &=\frac{\beta_c}{\mpl}\,\dot{\delta\phi},\\
 \dot{\bm v}_c+\Big(H+\frac{\beta_c\dot\phi}{\mpl}\Big)\bm v_c
 &=-\frac{1}{a}\bm\nabla\Psi-\frac{\beta_c}{a\mpl}\bm\nabla\delta\phi,
\end{align}
where $\bm\nabla$ is the comoving gradient; Eq.~\eqref{eq:eulerc} is the same
equation written with the physical gradient. The extra drag is
$\Gamma_c=\beta_c\dot\phi/\mpl=H\xi_c$. Baryons obey the same equations with
$\beta_c\to\beta_b^{\rm(em)}\to0$, Eq.~\eqref{eq:eulerb}. On sub-horizon scales,
$k\gg aH$, and neglecting time derivatives of $\delta\phi$ relative to spatial
ones, the scalar perturbation satisfies
\begin{equation}
 \big(k^2+a^2m_\phi^2\big)\delta\phi=-a^2\frac{\beta_c}{\mpl}\rho_c\delta_c,
\end{equation}
with $m_\phi^2$ as in Eq.~\eqref{eq:mphi}, while the Poisson equation is
$k^2\Psi=-4\pi G_Na^2(\rho_c\delta_c+\rho_b\delta_b)$ with
$4\pi G_N=(2\mpl^2)^{-1}$. Inserting $\delta\phi$ into the CDM Euler equation, the
total force on CDM from a CDM overdensity is the Newtonian one multiplied by
$1+2\beta_c^2k^2/(k^2+a^2m_\phi^2)$, which is Eq.~\eqref{eq:fifthforce}. The force
on baryons from any overdensity, and the force on CDM from a baryonic overdensity,
are Newtonian up to $\beta_b^{\rm(em)}$; lensing, which responds to $\Phi+\Psi$, is
likewise unmodified at this order. This is the sense in which the model is a
dark fifth force rather than a modification of gravity.

\emph{Local field.} With $\beta_b^{\rm(em)}\to0$ the only local source of $\phi$ is
dark matter. For the Galactic density $\rho_c\simeq0.4\,{\rm GeV\,cm^{-3}}$,
Eq.~\eqref{eq:mphi} gives $m_\phi\simeq\beta_c\times7\times10^{-40}\,{\rm GeV}$,
that is $m_\phi^{-1}\simeq 9\,{\rm Mpc}/\beta_c$, compared with
$m_\phi\simeq0.9\,\beta_cH_0$ from the cosmological density. The field is therefore
unscreened on every scale from the laboratory to the Galaxy, and its local time
derivative is the cosmological $\dot\phi$ up to corrections of order the local
density contrast times $(H_0/m_\phi)^{2}$, which are negligible.

\bibliography{clocks_dark_energy}

\end{document}